 \documentclass[onecollarge,natbib]{svjour2}

 \usepackage{graphicx}
\usepackage{epsfig}

\def\sla#1{\rlap\slash #1}

\newcommand{\be}{\begin{equation}}
\newcommand{\ee}{\end{equation}}
\newcommand{\cc}{\c c}

\newcommand{\bee}{\begin{eqnarray}}
\newcommand{\eee}{\end{eqnarray}}

\bibpunct{[}{]}{;}{n}{}{,} 
\smartqed  
\usepackage{graphicx,amsmath,amssymb}
\usepackage{color}
\usepackage{enumerate,array}

\usepackage{multicol}

\definecolor{navyblue}{rgb}{0.3,0.3,1}
\definecolor{purple}{rgb}{0.6,0,0.5}
\usepackage[colorlinks=true, pdfstartview=FitV, 
linkcolor=purple, citecolor= navyblue, urlcolor=navyblue]{hyperref}

\journalname{Few Body Systems}

\begin{document}

\title{ Electromagnetic spin-1 form-factor free of zero-modes 
\footnote{Presented by J.~P.~B.~C.~de Melo at LIGHT-CONE 2014,~May,~NCSU-USA}}  
\author{J.~P.~B.~C.~de~Melo$^1$,
Anac\'e N. da Silva$^1$, Clayton S. Mello$^{1,2}$\and T.~Frederico$^2$}
\institute{J.~P.~B.~C.~de~Melo and Anac\'e N. Silva \at
                Laboratorio de F\'isica 
Te\'orica e Computa\cc\~ao Cient\'ifica, 
Universidade Cruzeiro do Sul, 01506-000, S\~ao Paulo, Brazil$^1$. \\
                \and T.~Frederico and Clayton S. Mello \at
                Departamento de F\'isica, 
Instituto Tecnol\'ogico de Aeron\'autica, DCTA, 
12228-900, S\~ao Jos\'e dos Campos, Brazil$^2$.}

\date{Version of \today}

\maketitle

\begin{abstract}
The electromagnetic current~$J^+$ for spin-1, is used here to extract the electromagnetic form-factors of 
a light-front constituent quark model. The charge ($G_0$), magnetic  ($G_1$) and quadrupole  
$G_2$ form factors are calculated using different prescriptions 
known in the literature, for the combinations of the
four independent  matrix elements of the current between the polarisations states 
in the Drell-Yan frame.
However, the results for some  prescriptions  relying only on the valence 
contribution breaks the rotational symmetry 
as they violate the angular condition. In the present work, we use  
some relations between the matrix elements of the 
electromagnetic current  in order to eliminate the breaking of the rotational symmetry, 
by computing the zero-mode 
contributions to matrix elements resorting only to the valence ones.
\keywords{Spin-1 Particles  
\and Electromagnetic Current \and Electromagnetic Form Factors 
\and Light-Front Field Theory.}
\end{abstract}

\section{Introduction}
\label{intro}
Light-front models are useful to describe hadronic bound states,
like mesons or baryons due to its particular boost
properties~\cite{Terentev76,Terentev80,Brodsky98}. However, the light-front
description in a truncated Fock-space breaks the rotational
symmetry because the associated transformation is a dynamical
boost~\cite{Pacheco97,Pacheco98,Naus98,Pacheco992,Bruno2013,Bruno2014}. 
Therefore, an analysis with
covariant and analytical models, can be useful to pin down  the main
missing features in a truncated light-front Fock-space description
of the composite system. In this respect, the rotational symmetry
breaking of the plus component of the electromagnetic current, evaluated with
the Drell-Yan condition (momentum transfer $q^+=q^0+q^3=0$), was recently analysed relying on a model  of a composite spin 1 
two-fermion bound state with an analytical form of the vertex~\cite{Pacheco97,Pacheco98}.  
It was shown~\cite{Pacheco97,Pacheco98,Pacheco992,Bakker2002,Choi2004,Pacheco2003} that
if pair terms are ignored in the evaluation of the matrix elements
of the electromagnetic current, the covariance of the electromagnetic 
form factors is broken. The scalar nature of the form factors is restored 
only when pair terms or zero modes contributions are included in the computation of 
the current matrix elements~\cite{Pacheco98,Pacheco992,Choi2004,Pacheco99}. 

However, the extraction of the  electromagnetic form factors of a spin-1
composite particle from the microscopic matrix elements of the
plus component of the current ($J^+=J^0+J^3$) in the Drell-Yan
frame, based only on the valence component of the wave function, is plagued by 
ambiguities due to an expected rotational invariance breaking with a truncated state~\cite{Inna84,Inna89}. 
In the Breit frame, with the particular choice of the momentum transfer
along the transverse direction, where the Drell Yan-condition is valid, the plus component of the 
electromagnetic current has four independent  matrix elements, although only  three  form factors, 
$G_0,G_1$~and~$G_2$ are sufficient to parametrize the  current of 
 a vector particle. Then,  one is in principle free to choose different combinations of 
these matrix elements to compute form factors.  But, these matrix elements satisfy 
an identity,~the angular condition~\cite{Inna84,Inna89}, which is violated 
when only the valence contribution is considered~\cite{Pacheco97}. 
In the corresponding literature, different extraction schemes for 
evaluating the electromagnetic form factors for spin-1 particles were 
proposed in~\cite{Inna84,Hiller1992,Frankfurt93,CKP1988}. The difference between 
the various extraction schemes amounts to the choice of matrix element of  the current 
 eliminated by assuming  the angular condition. 

It was verified numerically, some time ago ~\cite{Pacheco97}, that the $\rho$-meson electromagnetic 
form factors obtained from an analytical and covariant model of the vertex 
using the prescription proposed in ~\cite{Inna84} with only the valence contribution,
 were in close agreement with the corresponding covariant  results. 
 It was demonstrated in ~\cite{Bakker2002} with a simplified model of the vertex and 
  light-cone polarizations to compute matrix elements, that the prescription from  ~\cite{Inna84} eliminates 
zero-modes contributions to the form factors.  Here, we will generalize such results by deriving 
analytical relations for the matrix elements~\cite{Pacheco2012}, where pair terms are implicitly included, 
even considering only the valence contribution, and after that the 
numerical calculations of the spin-1 form factors with  all prescriptions found in the literature agree. 

We should add that zero modes or pair terms in the matrix elements of the 
electromagnetic  current appear in different ways. For example, in the case 
of spin-0 particles, like for the pion or kaon,  
the plus component of the electromagnetic current  calculated in the 
Breit frame with the Drell-Yan condition does not have such 
contributions  in simple analytical and covariant models of 
the vertex~\cite{Pacheco99,Pacheco2002,Pereira2005,Otoniel2012,Kazuo2014}.

\section{Covariant Model  for  Composite Vector Particle}

The electromagnetic current of a spin 1 particle has the 
following general form: 
 \begin{equation}
 J_{\alpha \beta}^{\mu}=[F_1(q^2)g_{\alpha \beta} -F_2(q^2)
 \frac{q_{\alpha}q_{\beta}}{2 m_v^2}] (p^\mu + p^{\prime \mu})
  - F_3(q^2)
 (q_\alpha g_\beta^\mu- q_\beta g_\alpha^\mu) \ ,  
\label{eq:curr1}
 \end{equation}
 where $m_v$ is the mass of the vector particle, $q^\mu$ is the
 transfer momentum, $p^\mu$ and $p^{\prime  \mu}$ are the on-shell initial
 and final momentum, respectively.  Here, the spin-1 particle is identified with
 a $\rho$ meson.  Linear combinations of the covariant form factors
 $F_1$, $F_2$ and $F_3$ allow to get the charge ($G_0$), magnetic
 ($G_1$) and quadrupole ($G_2$) form factors (see e.g.~\cite{Pacheco97}).

In the impulse approximation, the matrix elements of the electromagnetic 
current ${\cal J}^+_{ji}$, are obtained from the loop integral~\cite{Pacheco97}, 
\begin{equation}
{\cal J}^+_{ji} =  \imath  \int\frac{d^4k}{(2\pi)^4}
 \frac{ Tr[\epsilon^{'\nu}_j \Gamma_{\beta}(k,k-p_f)
(\sla{k}-\sla{p_f} +m)  
\gamma^{+} 
(\sla{k}-\sla{p_i}+m) \epsilon^\mu_i \Gamma_{\alpha}(k,k-p_i)
(\sla{k}+m)]
\Lambda(k,p_f)\Lambda(k,p_i) }
{(k^2 - m^2+\imath \epsilon)  
((k-p_i)^2 - m^2+\imath\epsilon) 
((k-p_f)^2 - m^2+\imath \epsilon )}
\label{elmcu}
\end{equation}
where ${\epsilon^\prime_j}$ and ${\epsilon_i}$ are the
polarization four-vectors of the final and initial states,
respectively and $m$ is the quark mass. 
The electromagnetic form factors  are calculated in the Breit frame
with the Drell-Yan-West condition, $q^+=0$, 
which gives the  momentum  transfer
$q^\mu=(0,q_x,0,0)$, the particle initial momentum
$p^\mu=(p^0,-q_x/2,0,0)$ and the the final one
$p^{\prime\mu}=(p^0,q_x/2,0,0)$. 
We use the definition  $\eta=-q^2/4 m_{v}$, and then the energy of the vector particle is
$p^0=m_{v}\sqrt{1+\eta}$. 

The polarization four-vectors in the instant-form basis are given by 
$\epsilon^{\mu}_x =(-\sqrt{\eta},\sqrt{1+\eta},0,0),~~
\epsilon^{\mu}_y=(0,0,1,0),~~  \epsilon^{\mu}_z=(0,0,0,1),$ 
for the initial state and by, 
$\epsilon^{\prime \mu}_x=(\sqrt{\eta},\sqrt{1+\eta},0,0),~~
\epsilon^{\prime\mu}_y=\epsilon^{\mu}_y ,~~
\epsilon^{\prime\mu}_z=\epsilon^{\mu}_z,$ 
for the final state. 

In order to make finite the photon-absorption amplitude; 
we use following regularization function 
$\Lambda(k,p_{i(f)}) = N/((p-k)^2-m^2_R +\imath \epsilon)^2$ 
associated with a model for the  vector particle vertex with a regularization parameter $m_R$. 
The vertex function of the spin-1 particle ~\cite{Pacheco97} in terms of the quark momentum is 
given by
\begin{equation}
\Gamma^\mu (k,p) = \gamma^\mu -\frac{m_v}{2}
\frac{2 k^\mu -p^\mu}
{ p.k + m_{v} m -\imath \epsilon}   \ .
\label{rhov}
\end{equation}
In the equation above, the vector particle is on-mass shell; 
and $m_v$ is the vector bound state mass.

In the next section, the extraction schemes of the electromagnetic form factor 
from the plus component of the electromagnetic current 
and the angular condition are discussed.

\section{Angular Condition and Prescriptions to Compute Vector Particle Form Factors}

The angular condition for plus component of the electromagnetic 
current, Eq.~(\ref{elmcu}), in the Breit frame ($q^+=0$), 
is given by the equation below, with the light-front,~($I^{+}_{m'm}$)
and instant form,~($J^+_{ji}$), spin-basis,~\cite{Pacheco97,Keister91}:
\begin{equation}
\Delta(q^2)=(1+2 \eta) I^{+}_{11}+I^{+}_{1-1} - 
\sqrt{8 \eta} I^{+}_{10} -
I^{+}_{00} \ = \ (1 + \eta)(J^+_{yy}-J^+_{zz})=0  \ .
\label{eq:ang}
\end{equation} 
In the instant form spin basis, 
the angular condition assumes a simple form, namely, 
$J^{+}_{yy}=J^{+}_{zz}$~\cite{Frankfurt93}, 
as can be seen directly in Eq.~(\ref{eq:ang}). It also
allows the freedom to combine in different ways the matrix elements of the plus component of the
vector particle current to compute  electromagnetic form factors~\cite{Cardarelli95}.

The prescription adopted by Grach and Kondratyuk~\cite{Inna84} corresponds to
eliminate the matrix element $I^+_{00}$ in the light-front spin basis, from the calculation of 
 the electromagnetic form factors. In references \cite{Bakker2002,Choi2004,Pacheco2012,deMelo2003,deMelo2012}, 
it was explored the fact that the zero-mode contribute only to the matrix element $I^+_{00}$ in 
the light-front spin basis. In Ref. \cite{Inna89} the prescription from Grach and Kondratyuk was used 
to  calculate deuteron form factors, where $I^+_{00}$ is eliminated by the angular 
condition, and in this case the results for the 
electromagnetic form factors are free of the zero mode 
contribution~\cite{Choi2004,Pacheco2012,deMelo2003}.
However,  zero modes  are present in the matrix elements of the  current
in the case of the other prescriptions \cite{Hiller1992,Frankfurt93,CKP1988,Cardarelli95}, and consequently
the electromagnetic form factors are not free of such contributions (see the discussion in \cite{Pacheco2012}).

In Ref.~\cite{Pacheco2012}, it was derived  relations fulfilled by the zero-mode or Z-diagram contributions to
matrix elements of the plus component of the current in the instant-form basis:
\begin{equation}
J^{+Z}_{xx} +\eta \  J^{+Z}_{zz}=0 \ ,    J^{+Z}_{zx} +\sqrt{\eta} \
J^{+Z}_{zz}=0\ ~~\text{and}~~ J^{+Z}_{yy}=0~. 
\label{finalcurr} 
\end{equation}
From the angular condition (\ref{eq:ang}) and the fact that $J^{+}_{yy}$ does
not have zero-modes, i.e, $J^{+Z}_{yy}=0$; 
we can write the non-vanishing  matrix elements for the Z-diagrams 
in terms of the valence matrix elements, starting with  $J^{+Z}_{zz}=J^{+V}_{yy}- J^{+V}_{zz}$.
The final form of the matrix elements of the  current, where the zero-mode contributions are eliminated in favor of the valence ones 
through the use of (\ref{finalcurr})  are:
\begin{eqnarray} 
J^{+}_{xx}& = & J^{+V}_{xx}-\eta \left(J^{+V}_{yy}-J^{+V}_{zz}\right) 
\nonumber \\  
J^{+}_{zx} & = & J^{+V}_{zx}-\sqrt{\eta}\left(J^{+V}_{yy}-J^{+V}_{zz}\right) \nonumber \\
J^+_{zz}& = & J^{+V}_{yy}~, 
\label{relations1}
\end{eqnarray}
and also for light-front spin basis,~$I^+_{m^{\prime}m}$, is written below:
\begin{eqnarray}
I^{+Z}_{11}  =  0,~I^{+Z}_{10} =  0,~I^{+Z}_{1-1}=0~, 
\ I^{+Z}_{00} =   (1+\eta )J_{zz}^{+Z} \ . 
\label{relations2}
\end{eqnarray}
The equations above, make clear, in the case of the light-front spin basis, that only 
$I^{+Z}_{00}$, has a zero mode contribution~\cite{Pacheco2012,deMelo2012}, 
similar results are obtained also in the ref.~\cite{Bakker2002,Choi2004}. 
The prescription of Grach and Kondratyuk~\cite{Inna84} is written below in terms of 
instant form spin basis and also in the light-front spin basis:
\begin{eqnarray}
G_0^{GK}& = &\frac{1}{3}[(3-2 \eta) I^{+}_{11}+ 2 \sqrt{2 \eta} I^{+}_{10} 
+  I^{+}_{1-1}] 
=  \frac{1}{3}[J_{xx}^{+} +(2 -\eta) J_{yy}^{+} 
+ \eta  J_{zz}^{+}], \nonumber \\
G_1^{GK} & = & 2 [I^{+}_{11}-\frac{1}{ \sqrt{2 \eta}} I^{+}_{10}]
=J_{yy}^{+} -  J_{zz}^{+} - \frac{J_{zx}^{+}}{\sqrt{\eta}},
\nonumber \\ 
G_2^{GK}&=&\frac{2 \sqrt{2}}{3}[- \eta I^{+}_{11}+
\sqrt{2 \eta} I^{+}_{10} -  I^{+}_{1-1}] 
=  \frac{\sqrt{2}}{3}[J_{xx}^{+}-(1+\eta) J_{yy}^{+} 
+ \eta  J_{zz}^{+}] \ ,
\label{inna}
\end{eqnarray}
and with the use of the relations given by the Eq.(\ref{relations1}) 
and Eq.(\ref{relations2}),  zero-mode contributions are absent
due to fact that the matrix elements $I^+_{00}$ is excluded \cite{Pacheco2012}.
Other prescriptions  from the literature are given by~\cite{CKP1988},
\begin{eqnarray}
G_0^{CCKP} & = &
\frac{1}{3 (1 + \eta)}  \biggl[ (\frac{3}{2}-\eta)(I^+_{11}+ I^+_{00}) +
5 \sqrt{2 \eta} I^+_{10} + (2\eta- \frac{1}{2}) I^+_{1-1}  \biggr] =
\frac{1}{6}[2 J_{xx}^{+} + J_{yy}^{+} + 3 J_{zz}^{+}] 
\nonumber \\
G_1^{CCKP}&=&
\frac{1}{ (1 + \eta)} \biggr[I^+_{11}+ I^+_{00} - 
I^+_{1-1} - \frac{2 (1-\eta)}{\sqrt{2\eta}} I^+_{10}\biggl] 
= -\frac{J_{zx}^{+}}{\sqrt{\eta}}
\nonumber \\
G_2^{CCKP}& = &
\frac{\sqrt{2}}{3 (1 + \eta)}  \biggl[  -\eta I^+_{11}+ 2 \sqrt{2 \eta}I^+_{10} 
-\eta I^+_{00} + (\eta + 2) I^+_{1-1} \biggr]  =  
\frac{\sqrt{2}}{3} [J_{xx}^{+}-J_{yy}^{+}] \ .
\label{cckp}
\end{eqnarray}

The prescription from reference~\cite{Hiller1992} builds the form factors of the spin-1 particle as follows:
\begin{eqnarray}
G_0^{BH} & = & 
\frac{1}{3 (1 + 2 \eta)}  \biggl[ (3 - 2 \eta) I^+_{00}+ 8 \sqrt{2 \eta}I^+_{10} 
+  2 (2 \eta -1 )I^+_{1-1} \biggr]   \nonumber \\ 
& =& \frac{1}{3 (1+2 \eta)}\biggl [
J_{xx}^{+} (1+2 \eta)+ J_{yy}^{+}(2 \eta-1) 
+  J_{zz}^{+}(3+2 \eta) \biggr]  ~, 
\nonumber \\
G_1^{BH}&=& 
\frac{2}{(1 + 2 \eta)}  \biggl[ I^+_{00} - I^+_{1-1} 
+ \frac{(2 \eta -1 )}{\sqrt{2 \eta}} I^+_{10} \biggr] \nonumber \\
&=& \frac{1}{(1+2 \eta)} \biggl[ \frac{J_{zx}^{+}}{\sqrt{\eta}}
 (1+2 \eta)- J_{yy}^{+} +  J_{zz}^{+} \biggr]~,
\nonumber \\
G_2^{BH}&=&
\frac{2\sqrt{2}}{3 (1 + 2 \eta)}  \biggl[ \sqrt{2 \eta }I^+_{10} -\eta I^+_{00} 
-( \eta +1 ) I^+_{1-1} \biggr] \nonumber \\
&=&
\frac{ \sqrt{2}}{3 (1+2 \eta)}
\biggl[ J_{xx}^{+} (1+2 \eta)- J_{yy}^{+}(1+ \eta) - \eta J_{zz}^{+} 
\biggr]~. \label{bh}
\end{eqnarray} 
The prescription from reference \cite{Frankfurt93} gives the following from factors: 
\begin{eqnarray}
G_0^{FFS}& = & 
\frac{2}{3 (1 + 2 \eta)}  \biggl[ (2 \eta +3 ) I^+_{11} 
+ 2 \sqrt{2 \eta} I^+_{10} - \eta I^+_{00} + (2 \eta + 1) I^+_{1-1} 
\biggr]  
 =  \frac{1}{3}[ J_{xx}^{+} + 2 J_{yy}^{+} ]  ~, 
\nonumber \\
G_1^{FFS}&  = &  G_1^{CCKP}~,   \nonumber \\
G_2^{FFS} & = & G_2^{CCKP}~.  \label{ffs}
\end{eqnarray}  
In the expressions of the form factors given by prescriptions detailed above~\cite{Hiller1992,Frankfurt93,CKP1988}
$I^+_{00}$ is present, which implies  zero-modes contributions to the 
electromagnetic form factors.  In the next section, we present the calculations of 
the electromagnetic form factors with these prescriptions.

\begin{figure}[tbh!]
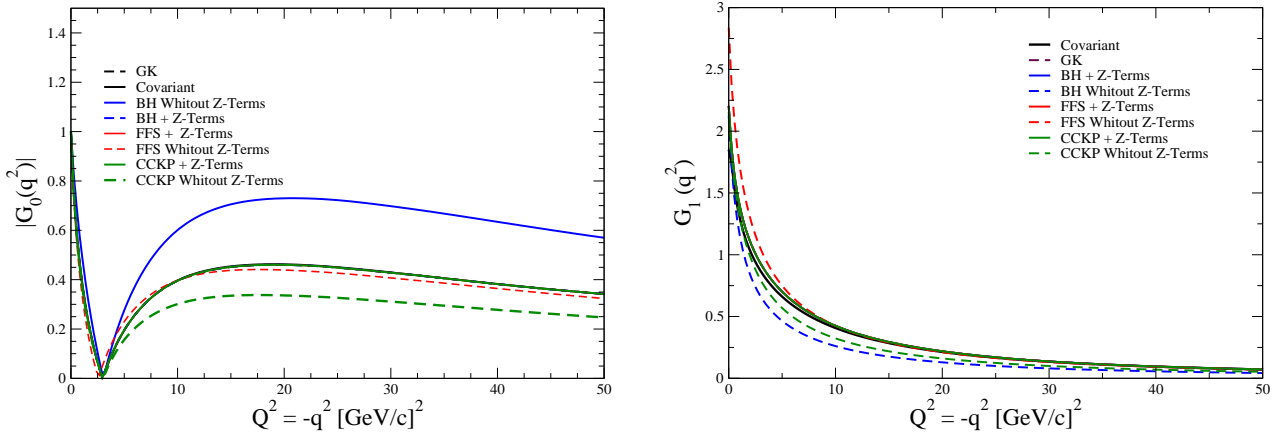

\centerline{
\epsfig{figure=fb1g0.eps,width=8.0cm} 
\hspace{0.4444000cm} 
\epsfig{figure=fb2g1.eps,width=8.0cm} 
} 
\caption{The figures show the electromagnetic form factors,
 $G_0$ and $G_1$, calculated with the plus component of the current 
with the  prescriptions labelled as GK\cite{Inna84} (\ref{inna}), 
BH \cite{Hiller1992} (\ref{bh}), CCKP\cite{CKP1988} (\ref{cckp}) and
FFS\cite{Frankfurt93} (\ref{ffs}) with and  without zero-mode contributions (Z-terms).}
\label{fig12}
\end{figure}

\begin{figure}[tbh!]
\vspace{0.6cm}
\centerline{
\epsfig{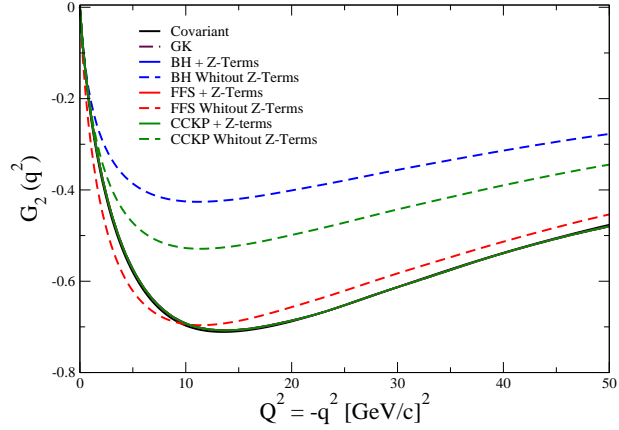}
 \hspace{1.00000004444000cm} 
}
\caption{The quadrupole form factor from calculations labelled as in Fig.~1.}
\label{fig34}
\end{figure}

\section{Results and Summary}

As suggested in Refs. \cite{Inna84,Inna89}, the elimination of the 
matrix elements $I^+_{00}$ through the angular condition,  as 
discussed in \cite{Choi2004,Pacheco2012}
cancels out the zero-modes contribution to the  form factors. 
In this case, it is possible to calculate the electromagnetic form factor 
 only from the valence terms of the matrix elements of the plus component of 
 the current in the Breit-frame with $q^+=0$.  
The relations (\ref{relations1}) and (\ref{relations2}) provides a practical 
way to compute the electromagnetic 
form factors of the spin-1 particle free of zero-modes, which is also 
verified numerically in the following figures 1, 2 and 3.  

The parameters for the model are the constituent quark mass,~$m=0.430~GeV$, 
$\rho$ meson mass,~$m_{\rho}=0.770~GeV$, 
and the regulator mass,~$m_{R}=~3.0~GeV$. 
The regulator mass is chosen to reproduce the experimental value of 
decay constant for the $\rho$ meson, which 
is $f_\rho=0.152 \pm 0.008 \ GeV$ \cite{PDG}. The present value of $m_R$ 
is different form the one used in \cite{Pacheco97},
where the decay constant was not fitted. 

Our new numerical results are shown in figures 1, 2 and 3. 
The calculations were performed for the charge, magnetic and quadrupole 
form factors,  with the instant form and the light-front spins basis.  
The prescriptions to extract form factors from the matrix elements of the plus
 component of the current given by Eqs. (\ref{inna}-\ref{ffs}) were compared 
 in the figures with the full covariant result (see ref. \cite{Pacheco97}).
 The charge form factor presents a zero, as shown in Fig. 1, which is dominated by 
 valence terms with the position weakly affected by the inclusion of 
 the zero-mode contribution to  the form factor. The zero modes present a 
 tail relevant for large momentum transfers for $G_0$ and $G_2$, while for $G_1$
 the effect is somewhat attenuated. These observations deserve further investigations 
 considering also different models for the $\rho$ Bethe-Salpeter amplitude, 
 as the symmetric model proposed in \cite{Pacheco2012}.
 
In summary,  in this work we checked numerically that after the inclusion of 
the contribution from zero-modes,  the covariance of the form factors is 
restored and all prescriptions adopted here  produced exactly the same results 
for the $\rho$ meson model adopted here. We also present a practical formula 
to calculate the matrix elements of the current in any spin basis 
computing only valence contributions.

\vspace{0.2cm}

{\bf Acknowledgments.} This work was supported in part by the
Brazilian agencies FAPESP (Funda\c{c}\~ao de Amparo \`a Pesquisa do
Estado de S\~ao Paulo), CNPq (Conselho Nacional de
Desenvolvimento Cient\'\i fico e Tecnol\'ogico) 
and CAPES (Coordena\c c\~ao de Aperfei\c coamento de Pessoal de N\'ivel Superior). 
Benefitted from access to the
computing facility of the Centro Nacional de 
Supercomputa\cc \~ao at the Federal University of Rio Grande do
Sul~(CESUSP/UFRGS).~J. P. B. C. de Melo thanks the organizer 
of the Light-Cone 2014 for the 
 invitation. 


\end{document}